\documentclass[a4paper,fleqn,usenatbib]{mnras}
\usepackage{txfonts}
\usepackage[T1]{fontenc}
\usepackage{ae,aecompl}
\usepackage{graphicx}    % Including figure files
\usepackage{color,ulem}
\usepackage{cancel}
\usepackage{slashed}
\usepackage{lipsum}
\usepackage{float}
\usepackage{relsize}
\usepackage{color}
\usepackage{xcolor}
\usepackage{multirow}

\def\dg{^\circ}

\newcommand{\Rm}{R_m} 
\newcommand{\Rf}{R_f} 
\newcommand{\Nm}{N_m} 
 
\newcommand{\zm}{z_m} 
\newcommand{\zf}{z_f} 
\newcommand{\tf}{t_f}

\def\perGpcCube{Gpc$^{-3}$}

\def\kmpersecperMpc{km s$^{-1}$ Mpc$^{-1}$}

\makeatletter

\newcommand{\Rmnum}[1]{\expandafter\@slowromancap\romannumeral #1@}
\makeatother

\title[Prompt emission vs afterglow detectability]{Detectability of Electromagnetic counterparts from Neutron Star mergers: prompt emission vs afterglow} 

\author[Mohan et al.]{
	Sreelekshmi Mohan,$^{1}$
		M. Saleem,$^{2}$
	and 
		Lekshmi Resmi$^{1,3}$\thanks{E-mail: l.resmi@iist.ac.in}
	\\
	$^{1}$Indian Institute of Space Science and Technology, Trivandrum, India.\\
	$^{2}$Chennai Mathematical Institute, Chennai, India. \\
	$^{3}$Anton Pannekoek Institute for Astronomy, Amdsterdam, The Netherlands.
}

\date{Accepted XXX. Received YYY; in original form ZZZ}

\pubyear{2019}

\begin{document}
	\label{firstpage}
	\pagerange{\pageref{firstpage}--\pageref{lastpage}}
	\maketitle
	
% Abstract of the paper
\begin{abstract} 
Electromagnetic observations of the first binary Neutron Star (BNS) merger GW170817 has established that relativistic jets can be successfully launched in BNS mergers. Typically, such jets produce emission in two phases:  $\gamma$-ray prompt emission and multi-wavelength afterglow. Due to relativistic de-boosting, the detectability of both these counterparts are dependent on the angle ($\theta_v$) between the observer's line of sight and the jet axis. We compare the detectability of prompt and afterglow emission from off-axis jets, assuming standard detector thresholds. We find that for top-hat jets, afterglow is a more potential counterpart than the prompt emission even with unfavourable afterglow parameters. For structured jets with a Gaussian profile, prompt emission is more promising than the afterglows at extreme viewing angles, under the assumption that the energy emitted in the prompt phase equals the kinetic energy of the outflow. Assuming a Gaussian jet profile, we forecast the population of $\gamma$-ray detections and find that extreme viewing angle events like GRB170817A will be rare. In our simulated sample, the observed isotropic equivalent energy in $\gamma$-rays is moderately correlated with the viewing angle, such that a low $E_{\rm iso, \gamma}$ is almost always associated with a high off-axis viewing angle.
\end{abstract}

%==================================================================================

%%%%%%%%%%%%%%%%%%%%%%%%%%%%%%%%%%%%%%%%%%%%%%%%%%
\section{Introduction}
Multi-messenger detection and observations of the first binary Neutron Star (BNS) merger by Gravitational Wave (GW) and Electro Magnetic (EM) observatories have revealed the immense potential of GW-EM synergetic studies \citep{MMA2017a}. The merger, GW170817 detected by AdvLIGO/Virgo, coincided temporally and spatially with the singularly faint short-hard burst GRB170817A detected by the Fermi Gamma-ray space telescope \citep{GRB2017a, Goldstein2017b}. The associated X-ray/radio non-thermal emission showed a distinct late onset and shallow rise compared to classical short GRBs and expectations of GRB afterglow models \citep{Hallinan2017b, Troja2017b, Kim2017, Mooley2017c}. The non-thermal transient could be explained as emission from an off-axis relativistic jet with a lateral structure in energy and bulk velocity \citep{Lazzati:2017zsj, Kathirgamaraju:2017igg, Margutti2017b, Resmi:2018wuc, Granot:2017gwa, Lyman2018a, DAvanzo:2018zyz}. However, a sub-relativistic quasi-spherical outflow with radial structure could also explain the non-thermal emission \citep{Hallinan2017b, Mooley2017c}. VLBI observations of the radio afterglow showed evidence for proper motion in the flux centroid and confirmed the presence of a relativistic off-axis jet emerging from GW170817 \citep{Mooley:2018dlz, Ghirlanda:2018uyx}. 

Both afterglow modelling studies \citep{Resmi:2018wuc, Granot:2017gwa, Lamb:2018qfn} and VLBI observations implied the viewing angle to be $\sim 20^{\circ} - 28^{\circ}$. Therefore, GRB170817A became the first GRB with a confirmed extreme off-axis viewing angle. Bounds on the angle between the observer's line of sight and the orbital angular momentum vector from GW observations also are in agreement with these numbers \citep{Mandel:2017fwk}.  

\cite{Resmi:2018wuc} used the posterior distributions of the kinetic energy, bulk Lorentz factor, viewing angle, and jet core angle from modelling multi-wavelength observations under a Gaussian structured jet model and showed that the observed isotropic equivalent energy in $\gamma$-rays could be reproduced. Several authors explained the observed properties of the prompt emission, such as the fluence and spectral peak, using a Gaussian structured jet model \citep{Meng:2018jtv, Ioka:2017nzl, Salafia:2019off} (however, see \cite{2019MNRAS.483.1247M, Matsumoto:2019sor} for a different conclusion). 
Motivated by the possibility of GRBs from structured jets associated with BNS mergers, several authors have also explored statistical properties of such bursts, particularly the population of GW-GRB joint detection, starting from an assumed luminosity function and redshift distribution \citep{Howell:2018nhu, Salafia:2019ldt}. 

Unlike other EM counterparts such as the kilonova, detections of the relativistic jet, if launched successfully, will be affected by the viewing angle. In this paper, we compare GRB prompt emission and afterglow detectability and find that the prompt emission is likely to be a more promising component from a structured relativistic jet. Following that, we study properties of the population of structured jets detected in $\gamma$-rays. The paper is organized in the following way. In section-2 we describe the calculation of the prompt emission fluence for both top-hat and Gaussian jets. In section-3 we compare afterglow flux from these jets for different ranges of physical parameters and compare the detectability of afterglows and prompt emission for a given jet structure. In section-4 we do a population synthesis for prompt emission from the Gaussian jets. We compare the parameter space of the detected population with both GRB170817A and cosmological short GRBs. We conclude with a summary of our findings in section-5.

We assume  the flat $\Lambda$-CDM cosmology with $(\Omega_M,\Omega_K, \Omega_\Lambda) = (0.3,0,0.7)$ and the Hubble constant $H_0 = 70$ \kmpersecperMpc throughout this work.
%{\textcolor{red}{make sure all works are cited}}
%-------------------------------------------------------------
\section{Prompt emission}
\label{sect2}

\begin{figure*}
	\begin{center}
		\includegraphics[scale=0.4]{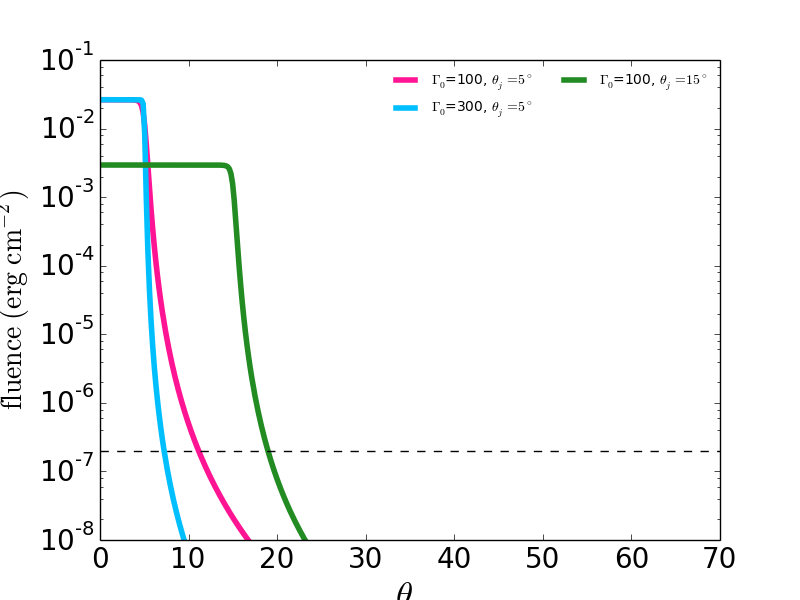} 
		\includegraphics[scale=0.4]{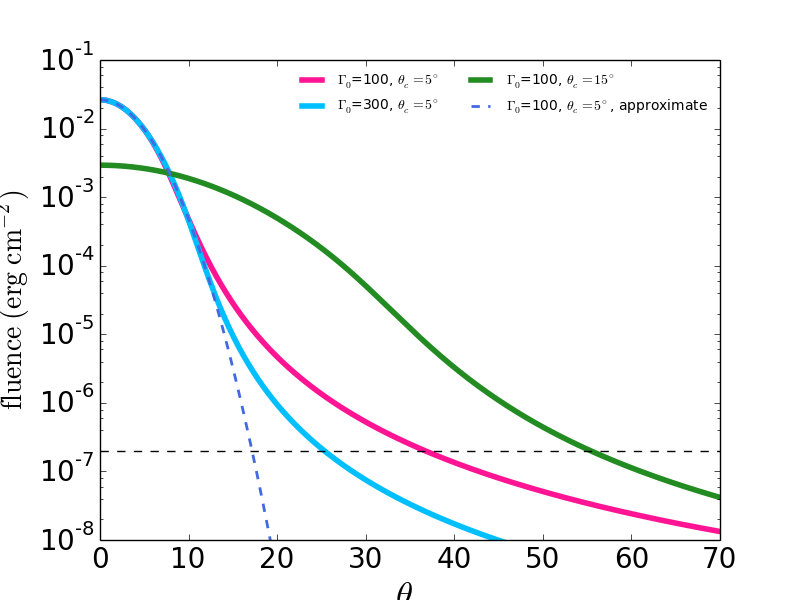} 
		\caption{Prompt emission fluence as a function of the observer's viewing angle $\theta_V$. $\theta=0$ corresponds to the jet axis. Two different jet models are considered: uniform top-hat model (left panel) and Gaussian structured jet model (right). Different curves correspond to different intrinsic jet parameters such as  the initial bulk Lorentz factor and the jet angle $\theta_j$ (or $\theta_c$). For the Gaussian jet of $\Gamma_c = 300$ and $\theta_c = 5^{\circ}$, we have also shown the behaviour of the approximate fluence (dashed blue curve, right panel) given by $\propto \epsilon(\theta)/d_L^2$, where  $\epsilon(\theta)$ is the energy per solid angle (see equation-\ref{eq4} in text). These curves are for a fixed value of energy emitted in $\gamma$-rays and luminosity distance ($E_{\rm tot,\gamma} = 10^{49}$ erg, $d_L=41$ Mpc). We have used a fluene threshold of $2 \times 10^{-7}$cgs (dashed line).}
		\label{fig-prompt}
	\end{center}
\end{figure*}
We calculate prompt emission fluence using purely energetic arguments following the expression derived by \cite{donaghy05} and \cite{Salafia:2015vla} using independent methods. The isotropic energy emitted in $\gamma$-rays, $E_{\rm iso, \gamma}$, as measured by an observer at a viewing angle $\theta_v$ with respect to the jet axis is,
\begin{equation}
E_{\rm iso, \gamma} (\theta_v)= \frac{E_{\rm tot,\gamma}}{4 \pi} \int d\Omega \frac{\epsilon(\theta)}{{\Gamma\!(\theta)}^4 \: \left[ 1-\beta\!(\theta) \cos{\theta_v} \right]^3}.
\label{eq1}
\end{equation}
 The jet is considered axially symmetric, and the the polar angle ($\theta$) measured from its axis is used to represent the angular profile in energy and bulk Lorentz factor.  $E_{\rm tot,\gamma}$ is the total energy emitted in $\gamma$-rays, $\epsilon(\theta)$ is a normalized function representing the energy structure of the jet, and $\Gamma(\theta)$ is the jet bulk Lorentz factor and $\beta(\theta)$ is the corresponding velocity. In this paper, we have considered only the forward moving jet, therefore, limits of $\theta$ integration is from $0$ to $\pi/2$ for Gaussian jets and $0$ to $\theta_j$ (half opening angle) for top-hat jets. $\epsilon(\theta)$ is normalized such that  $2 \pi \! \int \! d\cos{(\theta)} \: \epsilon(\theta) = 1$. For a homogeneous top-hat jet, obviously it is a step-function. The inferred total energy for an on-axis observer will be the same as $E_{\rm tot, \gamma}$ (see sections \ref{tp} and \ref{gj}). It needs to be noted that these calculations can only obtain the energetics integrated over the burst duration and spectrum. 

The same model is used in \cite{Kim2017} for top-hat jet and \cite{Resmi:2018wuc} for Gaussian structured jets to infer parameters of GRB170817A in combination with afterglow modelling.  There are also numerical calculations arriving at a similar profile of prompt emission fluence as a function of viewing angle for top-hat jets \citep{Woods:1999vx, Yamazaki:2002yb}. In addition, several authors have calculated the variation of prompt emission flux or fluence vs viewing angle \cite{LK17, Lazzati:2017zsj, Kathirgamaraju:2017igg, Ioka:2019jlj, Salafia:2019off}. 
%{\textcolor{red}{Is there a granot group paper which does the same?}}
%

%
\subsection{Top-hat jet}
\label{tp}
For a homogeneous top-hat jet, the normalized $\epsilon(\theta)$ will be $\epsilon(\theta) = \frac{1}{2\pi (1-\cos(\theta_j))}$ for $\theta \le \theta_j$ and $0$ otherwise, where $\theta_j$ is the jet half-opening angle. \citet{donaghy05} show that the integral in eq-\ref{eq1} is analytically solvable in  this case, leading to,
\begin{equation}
E_{\rm iso} (\theta_v)= \frac{E_{\rm tot,\gamma}}{2 \beta \Gamma^4 (1-\cos{\theta_j})}  \; \left[f(\beta -\cos{\theta_v}) - f(\beta \cos{\theta_j} -\cos{\theta_v}) \right],
\label{eq2}
\end{equation}
where $f(z) = \left[ (\Gamma^2(2 \Gamma^2-1)z^3 + (3 \Gamma^2 \sin^2{\theta_v} -1)z + 2 \cos{\theta_v} \sin^2{\theta_v} \right] /(z^2+\Gamma^{-2} \sin^2{\theta_v} )^{3/2}$.

For an on-axis observer ($\theta_v =0$), the term in square-brackets in eq-\ref{eq2} approaches  $ \Gamma^4$ for $\Gamma \gg 1\gg \cos{(\theta_j)}$, leading to the well known expression of $E_{\rm iso} (\theta_v =0) \sim  E_{\rm tot,\gamma}/(1-\cos{\theta_j})$.

Figure.~\ref{fig-prompt} (left) shows fluence profiles for the top-hat jet as a function of the viewing angle $\theta_v$ for a burst with $E_{\rm tot, \gamma} = 10^{49}$~erg at a distance of $41$~Mpc. We can see that on-axis fluence is not sensitive to the bulk Lorentz factor as mentioned above. As expected, the fluence has a steady value when $\theta \leq \theta_j$ but declines sharply afterwards.  At large $\theta_v$ (beyond what is shown in the figure), fluence falls as $(\Gamma \theta_v)^6$ as expected for a point source approximation. 
\subsection{Gaussian jet}
\label{gj}
For a Gaussian structured jet, $\epsilon(\theta) \propto exp{(-\theta^2/\theta_v^2)}$ \citep{Rossi:2001pk, Zhang:2001qt, LK17}. Following \citet{Resmi:2018wuc}, we assume
 for the bulk lorentz factor $\Gamma$, 
\begin{equation}
\Gamma_{\theta} \beta_{\theta} = \Gamma_0 \beta_0  \exp{-\left(\theta^2/2 \theta_c^2\right)}.
\label{eq3}
\end{equation}
The factor of $1/2$ in the argument is motivated by the assumption that both $\Gamma$ and the ejected mass $M_{\rm ej}$ follow the same angular profile leading to the resultant profile of the energy, $E = \Gamma M_{ej} c^2$. 

The normalized energy profile is,
\begin{equation}
\epsilon(\theta) = \frac{\exp{-(\theta^2/\theta_c^2)}}{\pi \theta_c^2 \left[1-\exp{\left(-\pi^2/4 \theta_c^2 \right)} \right] }.
\label{eq4}
\end{equation} 
To arrive at the normalized $\epsilon{(\theta)}$, we have used the approximate result $\int d\Omega \exp{-(\theta^2/\theta_c^2)} = \pi \theta_c^2 \left[1-\exp{-\left( \theta_j^2/\theta_c^2 \right)} \right]$. This requires approximating $\sin(\theta) \sim \theta$ in the expression of solid angle. As $\sin{(x)} \exp{-(x^2/a^2)} \rightarrow 0$ for $x \gg a$, major contribution to the integral is from small latitudes and this approximation is valid for moderate values of $\theta_c$. The fractional error in total energy by using this approximate expression ranges from $4 \times 10^{-4}$ at $\theta_c = 3\dg$ to $2 \times 10^{-2}$ at $\theta_c = 20\dg$.

In the calculation of $E_{\rm iso} (\theta_v)$ using equations-\ref{eq1}, \ref{eq3}, and \ref{eq4}, the integration over the azimuthal angle $\phi$ can be done analytically leading to,
\begin{equation}
E_{\rm iso} (\theta_v) = \pi E_{\rm tot,\gamma} \int_0^{\pi/2} d \theta \; \frac{(2 a^2 + b^2) \epsilon(\theta) \sin(\theta) }{\Gamma^4 (\theta) (a^2-b^2)^{5/2}}, 
\label{eq5}
\end{equation} 
where $a(\theta, \theta_v)= 1-\beta_{\theta}  \cos{\theta_v} \cos{\theta}$ and $b(\theta, \theta_v) = \beta_{\theta} \sin{\theta_v} \sin{\theta}$.  Throughout this paper, the half-opening angle of the Gaussian jet is assumed to be $\sim \pi /2$.  We continue the integration over the polar angle numerically to finally obtain $E_{\rm iso} (\theta_v)$. 

Similar to the top-hat jet, the Gaussian jet too results in an $E_{\rm iso} (\theta_v = 0) \sim E_{\rm tot,\gamma}/(1-\cos{\theta_c})$. This can be seen analytically and is explained in detail in the appendix. 

Figure.~\ref{fig-prompt} (right) shows fluence profiles for the Gaussian structured jet as a function of the viewing angle $\theta_v$ for a total $\gamma$-ray energy, $E_{\rm tot,\gamma} = 10^{49}$ ergs and the luminosity distance $d_L=41$ Mpc. The fluence has a smoothly declining profile compared to the top-hat jet case. On-axis fluence depends on $E_{\rm tot,\gamma}$ and $\theta_j$ (or $\theta_c$) alone, however the decline is sensitive to the bulk Lorentz factor of the outflow. 

We considered a threshold of $2 \times 10^{-7}$~erg/cm$^2$ for detection (see section-4). Under the top hat model, observers whose viewing angles are beyond a few degrees do not receive detectable emission whereas under the Gaussian jet model, the fluence values can be above the threshold up to much larger values of viewing angles. The highest viewing angle possible depends on jet energy, $\theta_c$, and $\Gamma$. The dashed curve in figure-\ref{fig-prompt} uses an approximate expression for the isotropic equivalent energy valid for smaller viewing angles \citep{Zhang:2001qt}, where the fluence is $\propto \epsilon(\theta)/d_L^2$. We see that the full integration over the jet surface increases the extent of viewing angles to which detection is possible. 
%Had \gwbns been a top-hat jet, at $\theta_v \sim 5 \times \theta_j$  \citep{Resmi:2018wuc}, prompt emission would not have been detectable 

\section{Afterglow}
To calculate  afterglow emission from top-hat and Gaussian jets, we have used the same model presented in \cite{Resmi:2018wuc}. Gaussian jet profile used in these calculations is given in equations \ref{eq3} and \ref{eq4}. We do not consider lateral expansion in the jet, which is more relevant in the late time flux decay (post the lightcurve peak) \citep{2018MNRAS.481.2581L}.  We calculated lightcurves for radio ($6$~GHz) and X-ray ( $5$~keV) frequencies. In addition to the kinetic energy $E_{\rm tot}$ in the outflow (which could be different from $E_{\rm iso, \gamma}$) and the jet opening angle (or core angle in case of Gaussian jet), the underlying synchrotron spectrum depends on the ambient density $n_0$, fraction $\epsilon_B$ and $\epsilon_e$ of shock generated thermal energy in downstream magnetic field  and in electrons respectively, index $p$ of the distribution of non-thermal electrons in energy space. The initial bulk Lorentz factor does not play a role in the observed flux once the fireball decelerates and enters the self-similar regime, which is valid  throughout the range of epochs where the flux is detectable. We have considered a range of values for ambient density $n_0$ and fractional energy content $\epsilon_B$ and assumed fixed values for other micro-physical parameters $\epsilon_e = 0.1$ (see \cite{Beniamini:2017jil} for a possibility of apparent universality of $\epsilon_e$ from radio observations) and $p=2.2$. All curves from figure-\ref{AGhat1} to \ref{AGgauss2b} are calculated for a luminosity distance of $d_L = 41$~Mpc. A viewing angle of $30\dg$, where the GW detection probability peaks \citep{Saleem:2017vly}, is used in all figures in addition to the highest viewing angle where prompt emission is detectable for a given set of jet parameters. For detection threshold, we used a somewhat conservative $0.05$~mJy for $6$~GHz lightcurves, and $2.5 \times 10^{-7}$~mJy for $5$~keV lightcurves. For X-rays, we chose this threshold based on the deepest detection of GRB170817 afterglow by \textit{Chandra}.

\subsection{Afterglow light curves in the top-hat jet model}
\begin{figure}
	\begin{center}
		\includegraphics[scale=0.3]{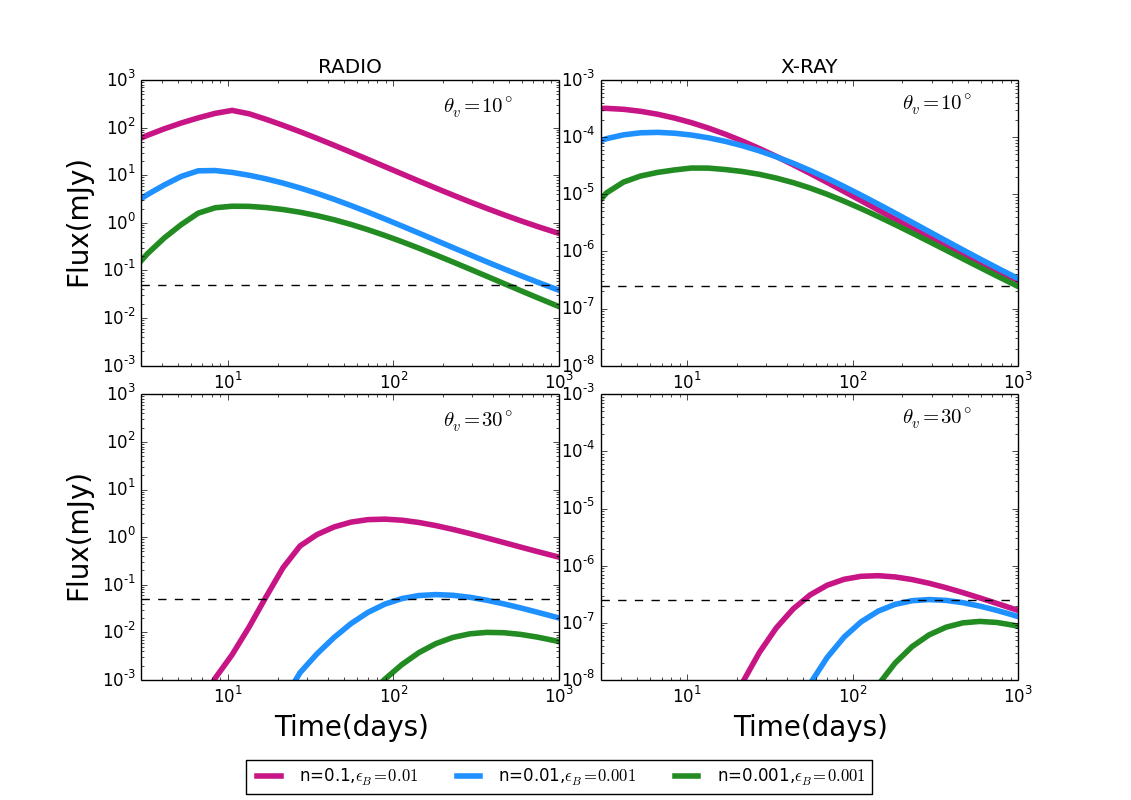} 
		\caption{Afterglow lightcurves in radio ($6$~GHz and X-ray ($5$keV) for the top-hat jet, for three different combinations of $n_0$ and $\epsilon_B$. Jet half-opening angle is $5^{\circ}$. Afterglow kinetic energy is assumed to be equal to the total energy content in prompt emission presented in figure-\ref{fig-prompt}, $10^{49}$~ergs. $d_L = 41$~Mpc, $\epsilon_e = 0.1$, and $p=2.2$. Prompt emission fluence from a  jet of same parameters ($E_{\rm tot}$ and $\theta_j$) goes below detectability limit for a $\theta_v \sim 10^{\circ}$. However, the afterglow emission remains well above the detectable limit even for low magnetic field and number density. For relatively large values of $n_0$ and $\epsilon_B$, afterglow is detectable at $\theta_v 30^{\circ}$.}
		\label{AGhat1}
	\end{center}
\end{figure}

\begin{figure}
	\begin{center}
		\includegraphics[scale=0.3]{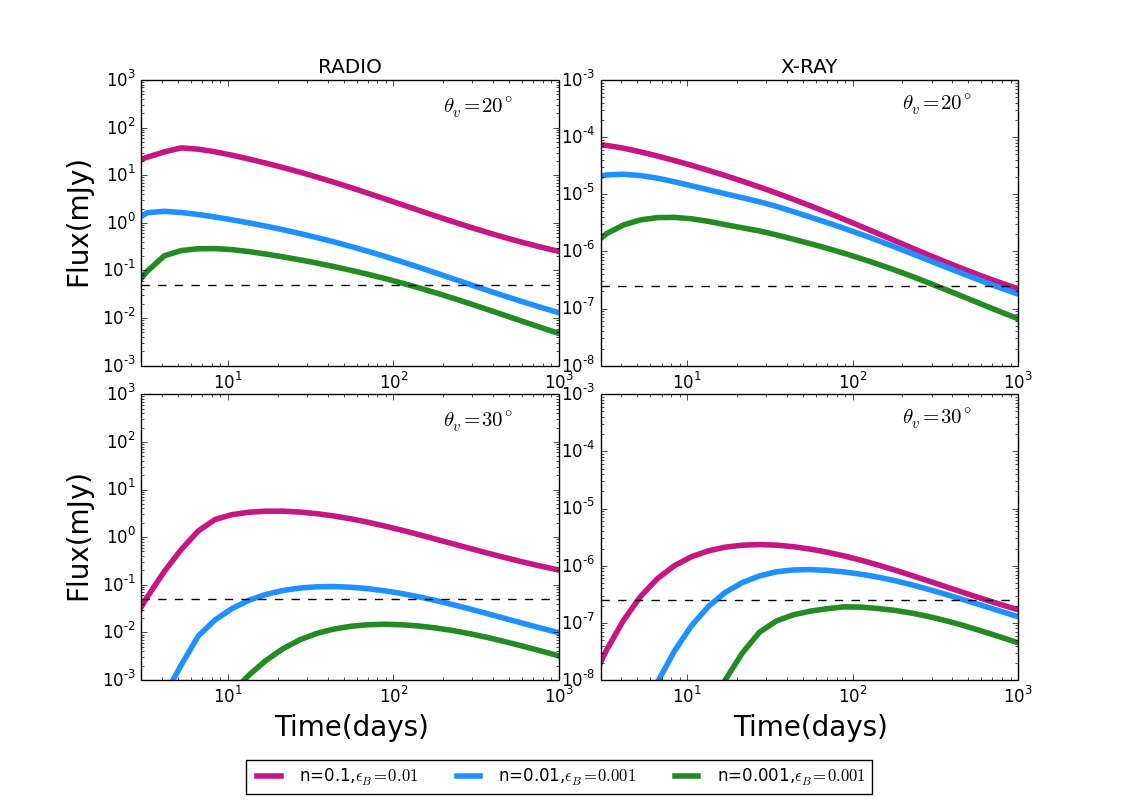} 
		\caption{Afterglow lightcurve for the same jet as in figure-\ref{AGhat1} but for $\theta_j = 15^{\circ}$. Since for prompt emission from a jet with initial bulk Lorentz factor $100$ (green curve in figure-\ref{fig-prompt} left panel), upto a $\theta_v$ of about $20^{\circ}$ the fluence is visible, we present afterglow flux for $\theta_v$ of $20^\circ$ and $30^\circ$. Even with a low $n_0$ and $\epsilon_B$, afterglow is well above detection limit for $\theta_v = 20^{\circ}$, and for high $n_0$ or $\epsilon_B$ at a larger viewing angles also.}
		\label{AGhat2}
	\end{center}
\end{figure}

We considered a jet with the same parameters used to estimate the prompt fluence in figure-\ref{fig-prompt}, such as $E_{\rm tot}=10^{49}$~ergs and opening angles $5$ and $15$ degrees. We have used three different combinations of ($n_0$, $\epsilon_B$), such as ($0.1,10^{-2}$), ($0.01,10^{-3}$), and ($0.001,10^{-3}$). In addition to lightcurves at $\theta_v = 30\dg$, for the jet of opening angle $5\dg$ ($15\dg$), we present lightcurves at a viewing angle of $10\dg$ ($20\dg$), where the prompt emission detectability nearly ends. See figures-\ref{AGhat1} and \ref{AGhat2}). 

We observe that while the prompt emission fluence goes below the detectability limit for a $\theta_v > \theta_j$ (see figure-\ref{fig-prompt}),  afterglow emission remains well above the detectable limit at large viewing angles even for low magnetic field and number density. Note that the outflow kinetic energy can be different from, and in most cases larger than the energy in prompt emission \citep{cenko2011}, resulting probably from the low efficiency of internal shocks \citep{Panaitescu:1999dw, Kumar:1999cv}. This will further improve the detectability of afterglows in comparison with prompt for larger viewing angles. 

\subsection{Afterglow light curves in the Gaussian structured jet model}
\begin{figure}
	\begin{center}
		\includegraphics[scale=0.3]{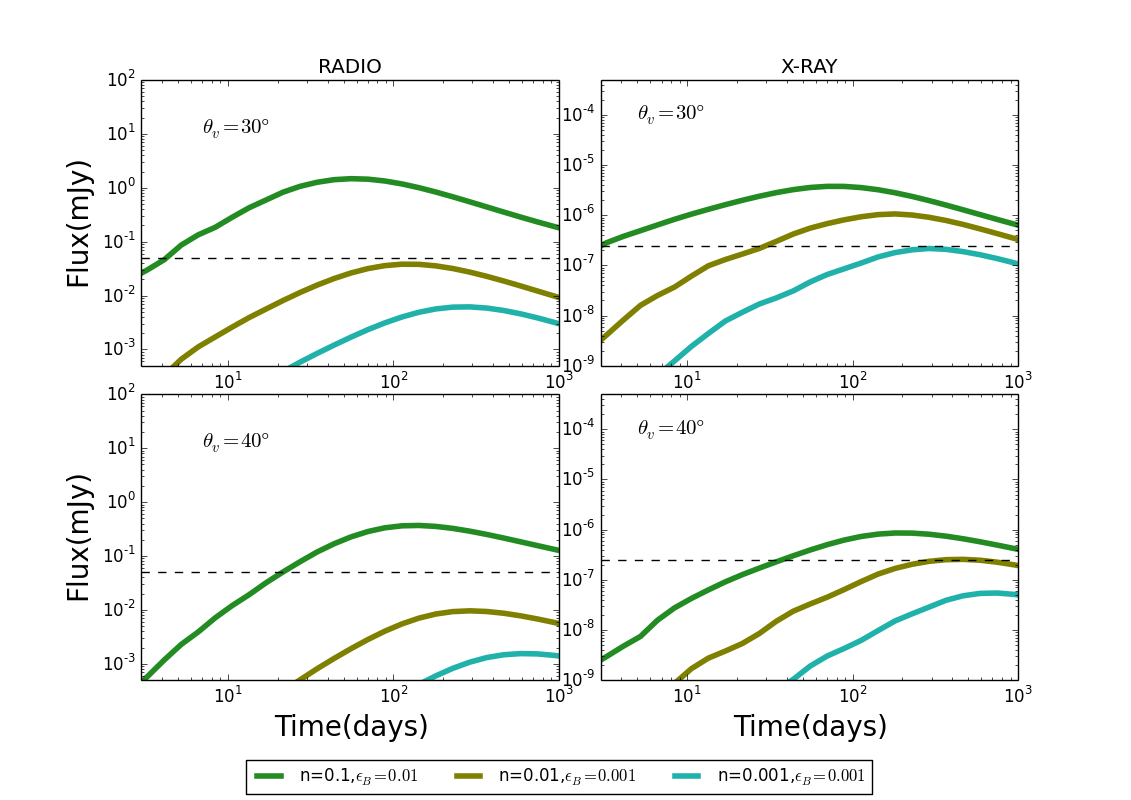} 
		\caption{Afterglow lightcurves for a Gaussian structured jet. The jet core $\theta_c = 5^{\circ}$, $E_{\rm tot,\gamma} = 10^{49}$~ergs, and $d_L = 41$~Mpc. While the prompt emission is detectable to large viewing angles as seen in figure-\ref{fig-prompt}, afterglow detectability at such angles is poorer unless the ambient medium is dense ($>0.1$~atom/cc) and/or the energy density in magnetic field is high.}
		\label{AGgauss1}
	\end{center}
\end{figure}

\begin{figure}
	\begin{center}
		\includegraphics[scale=0.3]{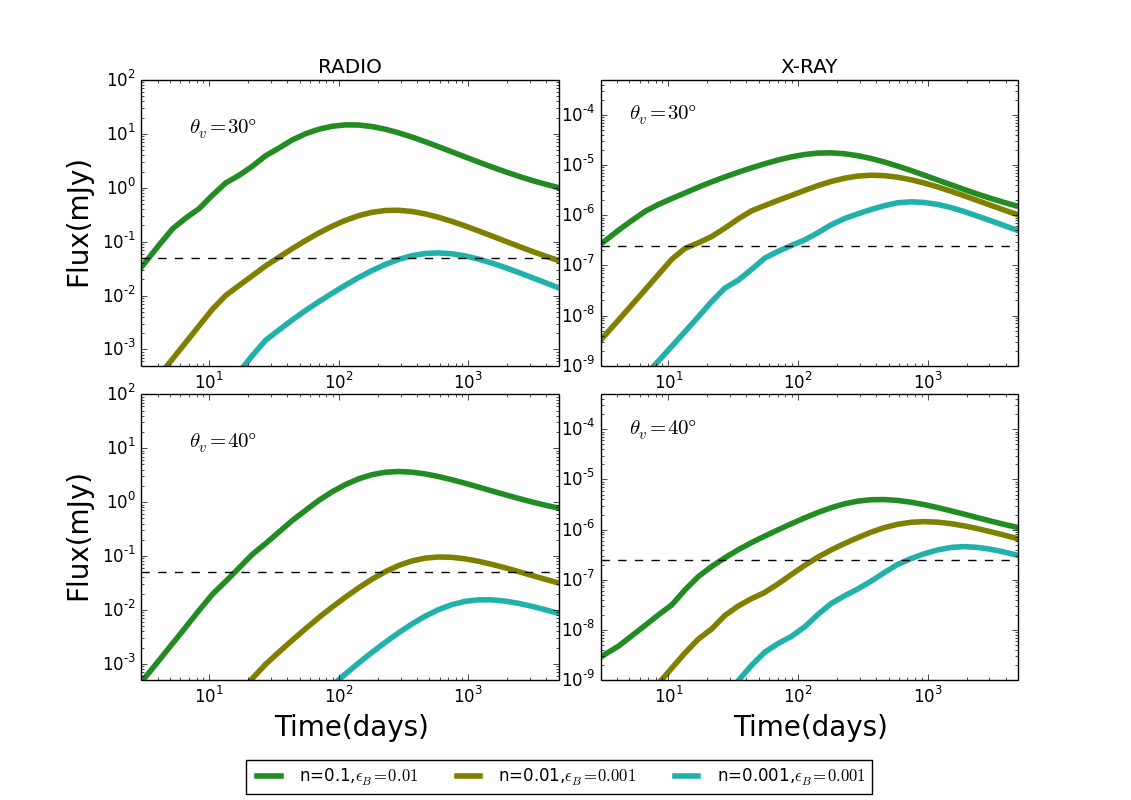} 
		\caption{Same as figure-\ref{AGgauss1} but for total kinetic energy of $E_T = 10^{50}$~ergs.}
		\label{AGgauss1b}
	\end{center}
\end{figure}

\begin{figure}
	\begin{center}
		\includegraphics[scale=0.3]{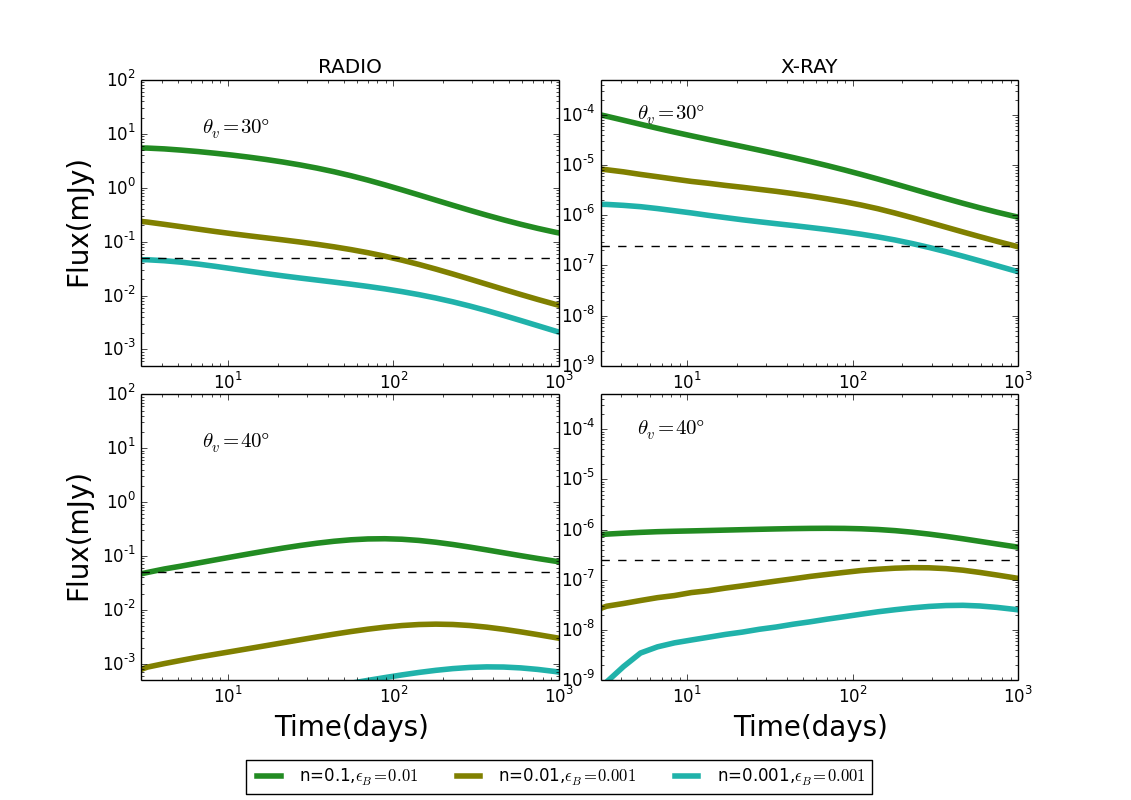} 
		\caption{Same as figure-\ref{AGgauss1} but for a jet of half-opening angle of $15^{\circ}$.}
		\label{AGgauss2}
	\end{center}
\end{figure}

\begin{figure}
	\begin{center}
		\includegraphics[scale=0.3]{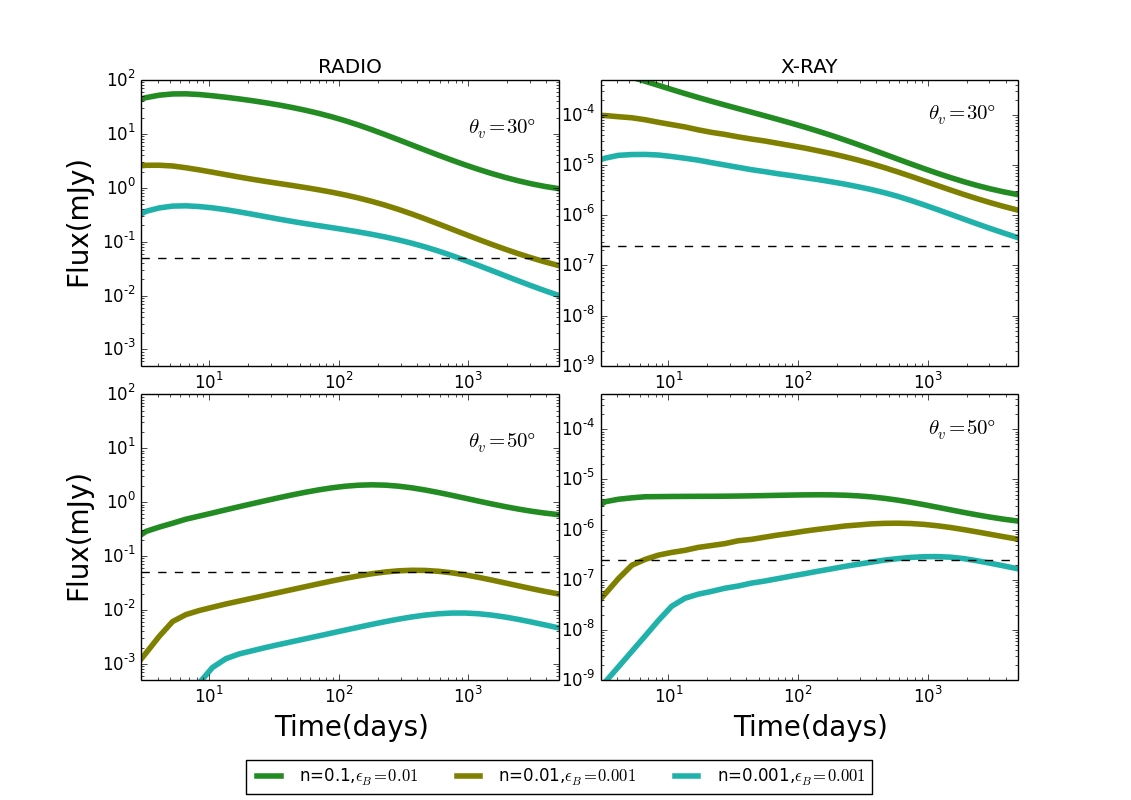} 
		\caption{Same as figure-\ref{AGgauss2} but for kinetic energy of $10^{50}$~ergs.}
		\label{AGgauss2b}
	\end{center}
\end{figure}

We have seen in section-\ref{gj} that prompt emission is detectable upto $\theta_v \sim 40\dg$ and $\sim 60\dg$ for $\theta_c$ of $5\dg$ and $15\dg$ respectively for the Gaussian jets with $E_{\rm tot,\gamma} = 10^{49}$~ergs and $d_L = 41$~Mpc. Therefore, in addition to $\theta_v = 30\dg$, we calculated afterglow lightcurves for $40\dg$ for the former case (figure-\ref{AGgauss1}) and for $50\dg$ for the latter case (figure-\ref{AGgauss2}). We see that the afterglow detectability is poor at these viewing angles, where prompt emission is still detectable, unless the ambient medium is dense ($>0.1$~atom/cc). A higher energy in the outflow obviously increases the flux, however improves the detectability only marginally for observers with large viewing angles (see figures \ref{AGgauss1b} and \ref{AGgauss2b} for lightcurves with $E_{\rm tot} = 10 E_{\rm tot, \gamma}$). All lightucurves assume $\epsilon_e = 0.1$, $p=2.2$, and $d_L = 41$~Mpc. 

From figures \ref{fig-prompt} to \ref{AGgauss2b}, we have shown that at a fixed distance, how the detectability of prompt emission and afterglow compare for different viewing angles, for the chosen detection thresholds. We see that prompt emission from Gaussian jets remain visible to observers at extreme viewing angles compared to that from top-hat jets. Therefore, when it comes to afterglows, where the peak flux is nearly similar for both jets with a given set of physical parameters, top-hat jets have more promise in afterglows than in prompt emission, while the Gaussian jets have more promise in prompt emission. However, it is important to note that viewing angles such as $40\dg$ and $50\dg$ are not highly probable for GW detected events. At $30\dg$ where the GW detection probability peaks, both prompt and afterglow are detectable for a Gaussian jet, like what has happened for GW170817 where the viewing angle is $24 - 32\dg$ according to VLBI results \citep{Mooley:2018dlz, Ghirlanda:2018uyx}. Therefore, for realistic cases, EM counterparts from Gaussian jets are detectable in both prompt and AG phase as opposed to those from top-hat jets where the detectability of prompt phase is poor at $30\dg$.

\begin{figure}
	\begin{center}
		\includegraphics[scale=0.3]{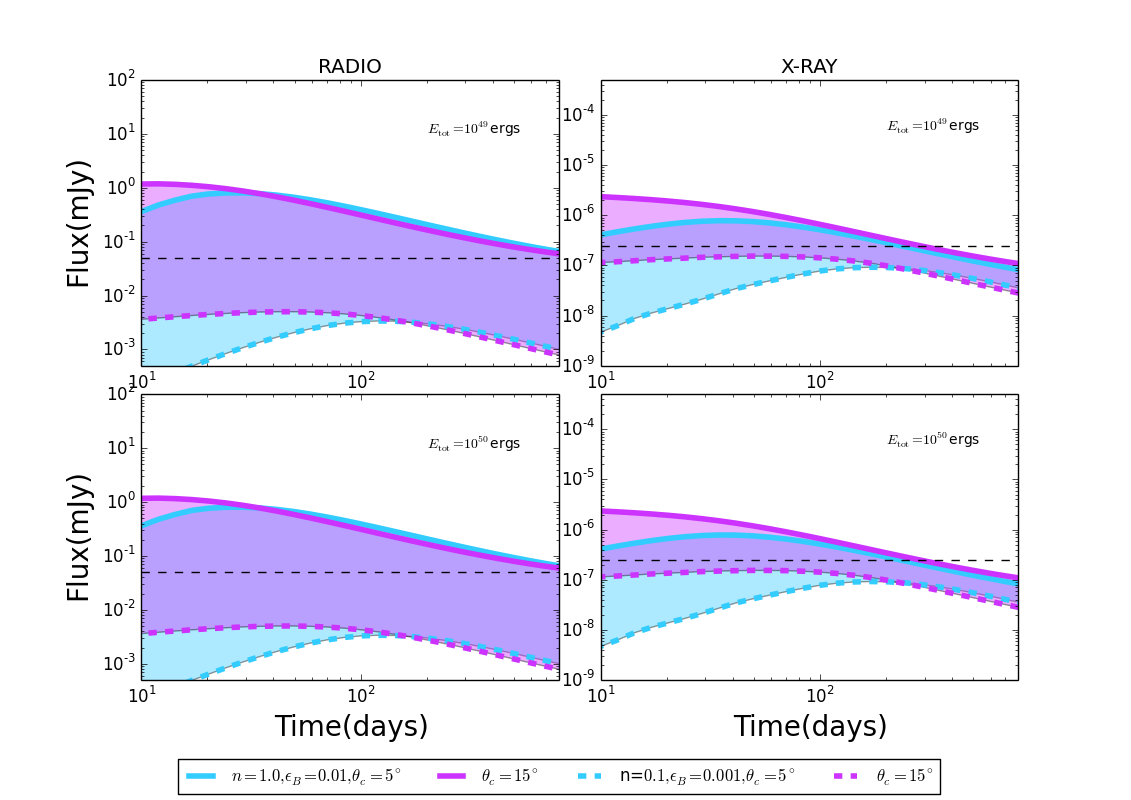} 
		\caption{Afterglow lightcurves for a viewing angle of $30\dg$ at $d_L = 150$~Mpc, close to the LIGO O3 BNS horizon. Solid curves correspond to $n_0 = 1.0$ and $\epsilon_B = 0.01$, while dashed curves are for $n_0 = 0.1$ and $\epsilon_B = 10^{-3}$. The shaded region represents lightcurves for $n_0$ and $\epsilon_B$ between these two  combinations. These are for Gaussian structured jets with $\theta_c = 5\dg$ (blue) and $\theta_c = 15\dg$ (violet). Top panel corresponds to a kinetic energy of $10^{49}$~ergs and the bottom panel is for an order of magnitude higher energy.  At the BNS horizon, a detectable afterglow is difficult for a viewing angle of $30\dg$.}
		\label{AGgaussfull}
	\end{center}
\end{figure}

Several authors have considered the detectability of afterglows with great detail before \citep{Duque:2019pfy, Gottlieb:2019vwb}, but for completeness we finally consider afterglows at a distance of $150$~Mpc, nearly at the BNS horizon for LIGO O3 and a viewing angle of $30\dg$. We plot the lightcurves for $E_{\rm tot} = 10^{49}$ and $10^{50}$~ergs and $\theta_c = 5\dg$ and $10\dg$  and see that afterglows are detectable if ambient density and downstream magnetic field are large (figure-\ref{AGgaussfull}).  Such high values of $n_0$ are unlikely if the merger is at the outskirts of the host galaxy as expected due to natal kicks. A wide range of $\epsilon_B$ is seen from modelling of long GRBs. Therefore the high values required for detection near the horizon is possible but may not be highly probable. 

As prompt emission is a promising counterpart for BNS mergers launching Gaussian jets, we next consider a population study of prompt emission and infer the parameter space allowing its detection. We do not consider joint GW-EM detections, instead, our aim is to study properties of GRBs associated with BNS mergers from a larger co-moving volume detected by $\gamma$-ray instruments such as \textit{Fermi} GBM and \textit{Swift} BAT.   
\section{A population study for the prompt-emission from structured jets}
Next, we perform a population study for prompt emission detections by \textit{Fermi} GBM under the Gaussian structured jet model, using the same method described in section-\ref{sect2} to estimate the observed fluence as a function of viewing angle. Since we calculate the isotropic equivalent energy (see equation-\ref{eq1}), we have used fluence as the detection criteria as opposed to the standard procedure where flux is estimated from luminosity function and used for detection condition. However, it must be noted that there is no well-defined fluence threshold for \textit{Fermi} GBM \citep{Bhat:2016odd}. Hence we considered the detection threshold to be $2 \times 10^{-7}$~erg cm$^{-2}$, where the fluence distribution for bursts with $T_{90} < 2$~sec peaks \citep{vonKienlin:2014nza, Bhat:2016odd}. About $25$\% of GBM short bursts have a lesser fluence,  therefore our results when it comes to bursts with low simulated fluence may be incomplete. %For example, this may result in a lower fraction of high $\theta_v$ detections than what could be possible in reality.

For the distribution of the jet parameters, such as $E_{\rm tot, \gamma}$, $\Gamma$, and $\theta_c$, we follow the below procedure. The prior on $E_{\rm tot, \gamma}$ is assumed to be a broken power-law ranging from $5 \times 10^{47} < (E_{\rm \gamma}/{\rm erg})) < 5 \times 10^{51}$, with a break at $10^{50}$~erg, and power-law indices $-0.53$ and $-3.4$ before and after the break energy respectively. We fixed the indices following \cite{Ghirlanda:2016ijf}, who fits the luminosity function of short GRBs. The ranges and break energy were fixed based on the agreement of the simulated population to the observed fluence distribution from \textit{Fermi} GBM. We choose uniform prior for the core angle ($5^{\circ}< \theta_c < 20^{\circ}$) and bulk Lorentz factor ($100 < \Gamma_0 < 500$) as no learned prior information is available for these parameters. Finally, we distributed the viewing angles uniformly in $\cos\theta_v$ between [0,1]. Before converging on the priors, we tested several other possibilities particularly for $E_{\rm tot, \gamma}$, such as different constant values, log-uniform distribution with different ranges, a single power-law with different indices and ranges, broken power-law with different parameters, etc.  In addition, we also attempted different ranges for $\theta_c$ and $\Gamma$ priors. The final distributions we used were able to broadly reproduce the $10-1000$~keV fluence distribution from the \textit{Fermi} 4-year catalogue \citep{Gruber:2014iza} for bursts with $T_{\rm 90} < 2$~sec.  

In order to decide the distribution in $d_L$, we took aid of the local BNS rate from AdvLIGO/Virgo \citep{gwtc-1}, the star formation history of the universe \citep{Madau:2014bja}, and a simple $1/\tau$ distribution for the delay-time ($\tau$) between the formation of a binary star system and its eventual merger as two neutron stars \citep{Guetta:2004fc}. The rate density of BNS mergers, which is a convolution of the cosmic star formation rate $\Rf$ and the delay time distribution $p\left(\tau\right)$ can be written as,
\begin{equation}
r_m (\zm) = \int_{\zm}^{\infty} \Rf(\zf) \, p\left(\tau(\zf,\zm)\right) \, \frac{d\tf}{d\zf} \, d\zf,
\label{eq-rz}
\end{equation}
where $\zm$ and $\zf$ are the redshifts corresponding to the epochs at which the BNS merger and binary formation happen respectively and the delay-time follows a power-law distribution given by, $p(\tau)\propto 1/\tau$, {with $\tau>\tau_{\rm min} = 10$~Myr. The Normalized rate density $\Rm$ can be written as,
\begin{equation}
\Rm(\zm) = \frac{\Rm(0)}{r_m(0)}\, r_m(\zm).
\label{eq-Rm}
\end{equation}
where $\Rm(0)$ is the local rate density which we assume to be the same as the BNS rates estimated from LIGO/Virgo detections~\citep{gwtc-1},  \textit{ie,} $\Rm(0) = 662^{+1609}_{-565}$ \perGpcCube. Given that, the cumulative number of mergers per year up to a redshift $\zm$ is given by, 
\begin{equation}
	\Nm(\zm) =  \int_{0}^{\zm}\frac{\Rm(z)}{1+z} \, \frac{dV_c(z)}{dz} \, dz,
	\label{eq-Nz}
\end{equation}	
where the factor $(1+z)$ in the denominator accounts for the cosmological time dilation and $\frac{dV_c}{dz}$ is the  differential co-moving volume element at $z$. 

We simulate a population as per Eq.~\ref{eq-Rm} (considering $\Rm(0) = 662$, the median value) and assume all of them to produce Gaussian structured jets.

\subsection{Detected population}
We computed prompt emission fluence for the above population using Eq.~\ref{eq5}, after applying bolometric and k-corrections. To estimate the correction factors, we used a Band function model for the prompt spectrum with indices $\alpha = -1.0$, $\beta = -2.2$, and $E_{\rm peak} = 800$~keV (we also estimated corrections for Comptonized spectrum and found that the difference between spectral models is negligible for our purpose).  For bolometric correction, following \cite{Howell:2018nhu}, we obtained the fraction of fluence between $10-1000$~keV and $1 - 10^4$~keV.  k-correction is estimated using the normal procedure (see for example \cite{WP15}). We found that the corrections never exceed a factor of $10$.

We find the rate of detection to be $521$ bursts per year, for fluence $> 2\times 10^{-7}$~erg/cm$^{2}$. Compared to the GBM sample of $T_{90} < 2$~sec bursts above the same fluence cut-off, this rate is larger by a factor of $\sim 18$. Accommodating for the sky coverage of {\textsl Fermi} GBM will reduce this excess by about a factor of $2$. The remaining excess can be well accommodated within the uncertainity in local merger rate, SFR models, and delay time distribution models. 

If all short GRBs emerge from Gaussian jets, the population of bursts detected by \textit{Fermi} so far correspond to a broad range of viewing angles. However, as expected for $\gamma$-ray detected events, a large fraction of the detections have $\theta_v < \theta_c$. The detected population has 64\% of the sources with $\theta_v / \theta_c  <1$  while the remaining 36\% are with $\theta_v > \theta_c$,  while the prior population contained 97\% with $\theta_v/\theta_c>1$. 

Though our simulations used \textit{Fermi} GBM detection threshold, we also compared the resulting $z$-distribution of the final detected sample with the observed one, which is almost entirely detected by \textit{Swift} BAT \citep{Fong:2015oha}. We obtained a good agreement between the observed and the simulated population, except that we obtain $5$\% of events to be at a $z>2.6$, the highest redshift for a short GRB so far \citep{2009GCN..9264....1L}.  This discrepancy could be because of a selection bias in the observed sample making optical afterglow detection or host galaxy identification difficult at large $z$. In addition, the difference in bolometric and k-corrections for GBM ($10-1000$~keV) and BAT ($15-150$~keV) could also have contributed to this difference.

Fig.~\ref{fig:rarity} shows the distribution of $d_L$ vs $\theta_v/\theta_c$ of the detected bursts from our simulation, color coded by observed fluence. We see that $2.5$\% of our detected sample contains bursts with fluence $> 10^{-5}$~erg/cm$^2$, while no bursts are present in the GBM sample above this level. Half of these bursts are at $d_L < 100$~Mpc, indicating a local population which is likely missing from the observed sample consisting of several years of data. This could mean that not all BNS mergers are successful in producing jets. %,
We overplot GRB170817a in this figure using parameters consistent with those inferred through afterglow modelling. We used a fluence of $1.4 \times 10^{-7}$~erg/cm$^{-2}$ \citep{Goldstein2017b}, $d_L = 41.3$~Mpc,  $\theta_c$ to be $5\dg$  and $\theta_v$ to be $20\dg$ \cite{Lamb:2018qfn}. This indeed shows that GRB170817A is a rare detection both in terms of its $d_L$ and $\theta_v/\theta_c$. One caveat in this conclusion is that we may be underestimating the fraction of faint events, due to our fluence cut-off of $2 \times 10^{-7}$~erg/cm$^2$ which is higher than the lowest detected \textit{GBM} fluence. It has to be noted that the rarity does not imply an inability of the structured jet model to reproduce GRB170817A energetics, but only indicates that such events are rare as a population. 

The distribution also hints that a small fraction of {\textsl Fermi} bursts may contain nearby events viewed at large angles. There have been claims of such local events from archival \textit{Fermi} and \textit{Swift} data \citep{Burgess:2017gwf, 2018NatCo...9.4089T, vonKienlin:2019otj}.  

For the case of joint detectability of the gravitational waves from the BNS merger and the prompt $\gamma$-rays, the detections will be a further subset of what is shown in Fig.~\ref{fig:rarity} and the possible space of $d_L$ and $\theta_v/\theta_c$ can be better constrained. We have not considered that in this simulations though the progenitors are all assumed to be BNS mergers. 

\subsection{Implications to short GRB observations}
One of the motivations behind the original hypothesis of a Gaussian structured jet for GRBs \citep{Rossi:2001pk, Zhang:2001qt} was a potential clustering seen in the kinetic energy of GRB fireballs \citep{Frail:2001qp}. Similarly, if the total energy emitted in GRB prompt emission is a constant, and if all the jets have characteristic core angles, a tight correlation of the isotropic equivalent energy with the viewing angle would result. However, from our simulations, we still find a correlation between these two quantities even when $E_{\rm tot, \gamma}$, $\theta_c$, and $\Gamma_c$ are drawn from broad distributions we have considered here (see figure-\ref{fig:eisocorr}). For the population we have simulated, the Spearman's rank correlation coefficient between the two parameters is $-0.36$.  However, observationally it is challenging to obtain $\theta_v$ for short GRBs, particularly at cosmological distances due to their faint afterglows. Therefore, $\theta_v$ information is more likely to emerge only for GW detected events, where the value is likely to remain in a narrow range around $20\dg - 40\dg$ \citep{Saleem:2017vly}. Hence, establishing this correlation will require deeper and longer follow-ups of short GRBs, likely possible in the era of the Square Kilometer Array. 

Nevertheless, in our simulations, the lowest viewing angle for bursts with $E_{\rm iso} < 10^{49}$~erg is $6^{\circ}$ (see figure-\ref{fig:eisocorr}), indicating that such low  $E_{\rm iso}$ values almost always indicate a large viewing angle. Motivated by this, we explored the short GRB population with $z$ information for low $E_{\rm iso}$ events \citep{Fong:2015oha}. Almost all bursts with $z$ information, even if from emission lines in the spectra of a putative host galaxy, are observed by \textit{Swift} BAT, while our population is estimated for \textit{Fermi} GBM. Therefore, we corrected the observed fluence of these bursts for the variation in $T_{90}$ expected between detectors,  the difference in detector bands, and $k(z)$. We found the isotropic equivalent $\gamma$-ray energy of GRB150101B, GRB050509B, GRB080905A, GRB060502B, and GRB061201 to be $\le 10^{49}$ ergs. Of these, GRB150101B is already discussed in the literature to be a potential off-axis event based on prompt and afterglow properties \citep{2018NatCo...9.4089T}. X-ray afterglows of both GRB050509B and GRB060502B are faint and are consistent with predictions of an off-axis outflow. The remaining two bursts have bright early X-ray emission, however it is not conclusive whether they are external forward shock origin or are from internal processes. In future, low $E_{\rm iso, \gamma}$ short GRBs could be targeted particularly for late and deep radio observations to probe their off-axis nature. 

\begin{figure}
	\includegraphics[scale=0.4]{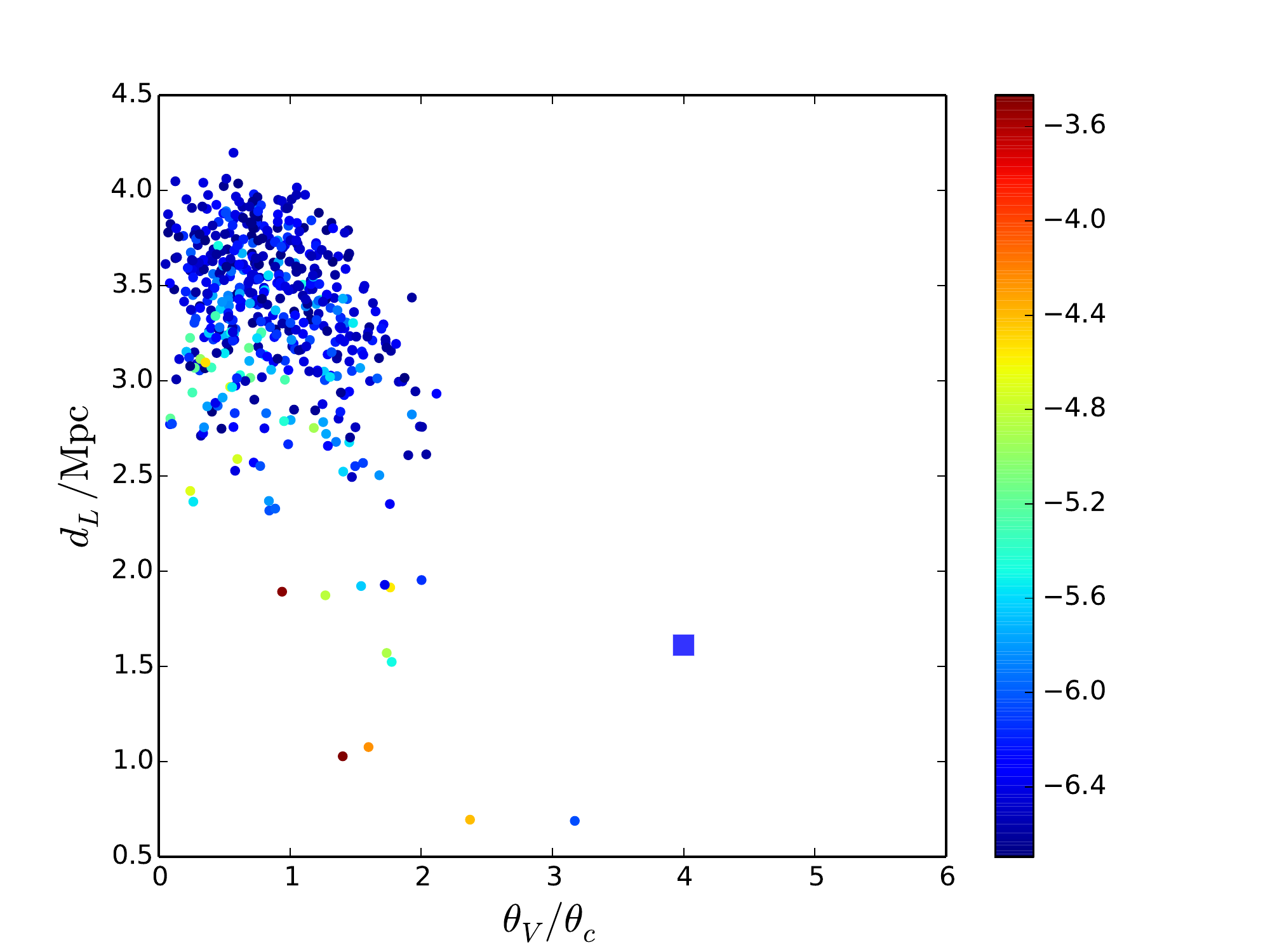}
		\caption{ Distribution of $d_L$ and $\theta_v/\theta_c$ for detected bursts from our simulation, color coded by fluence (in cgs units). Also plotted is GRB170817A (blue square) color coded by the observed GBM fluence. See text for the parameters of GRB170817A used in this figure. This shows $\gamma$-ray triggered short bursts cover a large range of viewing angles, with some local events likely seen at large viewing angles. Though our detection criterion underestimates low fluence events, GRB170817 is definitely a rare event as far its $d_L$ and $\theta_v/\theta_c$ are concerned. We see a few high fluence ($>10^{-5}$) events in our simulated sample which are not present in the Fermi detected sample. Most of these events are at the local universe. We also see that a part of the population, predominantly having low fluence, extends to larger distances than what is observed. This could be a selection effect against $z$ determination at large $z$. It must also be noted that short bursts with redshifts are almost always BAT bursts while this simulation is for \textit{Fermi} bursts. GBM may be in fact picking up more distant events than BAT due to its band extending to higher energies.}
	\label{fig:rarity}
\end{figure}
\begin{figure}
	\includegraphics[scale=0.4]{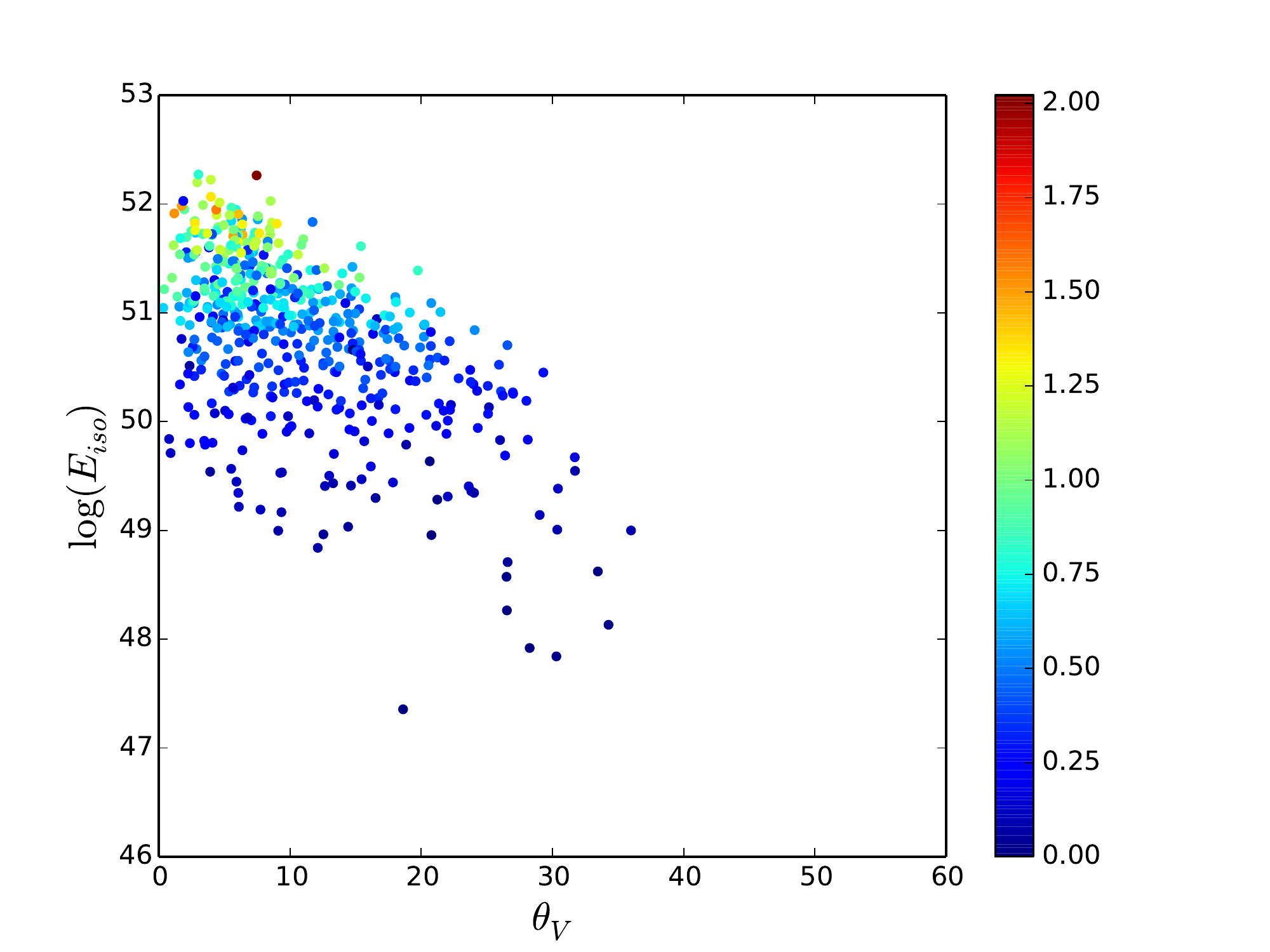}
	\caption{We see a correlation between $E_{\rm iso}$ and $\theta_v$ even when $\Gamma$, $\theta_c$ and $E_{\rm tot,\gamma}$ are drawn from a broad distribution. In this figure, we have limited our samples for $z < 2.6$ (color coded). However, measuring $\theta_v$ for EM detected bursts, particularly at large $d_L$, is a challenge. On the other hand, this correlation implies that bursts with very low $E_{\rm iso, \gamma}$ values are almost always observed at large viewing angles. Observed $E_{\rm iso, \gamma}$ does not go above $10^{51}$ for \textit{Swift} BAT bursts. However large $E_{\rm iso, \gamma}$ events in this sample are almost always at high redshift. Hence, most likely these events are present in the \textit{Swift} or \textit{Fermi} sample without having an estimate of the redshift due to selection biases.}
	\label{fig:eisocorr}
\end{figure}

\section{Summary and Conclusions}
In this paper, we have compared the detectability of electromagnetic counterparts to BNS mergers from relativistic jets. For jets with a top-hat structure, prompt emission is not expected to be detected at extreme viewing angles. However, for Gaussian structured jets like the one associated with GW170817, prompt emission can be detected to extreme off-axis angles ($50^{\circ} - 60^{\circ}$) at a distance of $41$~Mpc. For a range of afterglow parameters, such as the ambient medium density and magnetic field density, the afterglow in X-ray/radio is detectable for moderate viewing angles ($\sim 30^{\circ}$) for both type of jets at $41$~Mpc. However, for larger distances, high $n_0$ and $\epsilon_B$ are required to detect the afterglow at moderate viewing angles, while the prompt emission is still detectable if the jet has a Gaussian structure. 

We have seen that under a Gaussian structured jet model, prompt emission is a promising EM counterpart to the BNS mergers, especially at moderate to large viewing angles. Therefore, we undertook a population study of $\gamma$-ray prompt emission detections from structured jets. Our model only consider relativistic corrections across the jet surface to obtain a time and frequency integrated quantity, the measured fluence. Spectral and temporal variations due to viewing angle can not be accommodated in this model. Moreover, instead of the standard approach where peak flux is used for detection, we used fluence as a criteria for detection, which can modify the results in the context of faint events as a strict fluence threshold does not exist for \textit{Fermi} GBM. 

We found that events like GRB170817A at the nearby universe, having extreme viewing angles ($\theta_v \sim 4 \times \theta_c$) are rare. However, we find that $\sim 36$\% of the total detectable events by \textit{Fermi} GBM have $\theta_v > \theta_c$. Such bursts could be present in the current \textit{Fermi} and \textit{Swift} sample of short GRBs. A more realistic detection threshold can modify these results, particularly by helping the detection of faint, high $\theta_v/\theta_c$ events.  Our simulations also show that low $E_{\rm iso, \gamma} \le 10^{49}$ is almost always associated with high off-axis viewing angle.
%\bibliographystyle{mnras}
%\bibliography{ref} 
\input{mnras_draftv04.bbl}

%%%%%%%%%%%%%%%%%%%%%%%%%%%%%%%%%%%%%%%%%%%%%%%%%%

%%%%%%%%%%%%%%%%% APPENDICES %%%%%%%%%%%%%%%%%%%%%

\appendix
\section{Prompt fluence from Gaussian jet for an on-axis observer}

In this section, we show that for the Gaussian jet $E_{\rm iso} (\theta_v = 0) \sim E_{\rm tot,\gamma}/(1-\cos{\theta_c})$ for an on-axis observer. Eqn-\ref{eq5} gives the general expression for the Gaussian jet, before the final step which is the integration over the polar angle. Factors $a$ and $b$ in this equation  reduces to $a = 1-\beta_0 \cos{\theta}$  and $b=0$ for an on-axis observer, where $\beta_0$ is the velocity along jet axis, corresponding to $\Gamma_0$. Therefore,
\begin{equation}
E_{\rm iso} (0) = \frac{E_{\rm tot,\gamma}}{\theta_c^2} \int_0^{\theta_j} d \theta \frac{\sin{(\theta)}}{(1-\beta(\theta) \cos{(\theta)}^3} \frac{\exp^{-\theta^2/\theta_c^2}}{\left[\Gamma(\theta)^2\beta(\theta)^2 + 1\right]^{1/2}}.
\end{equation}
We have ignored the term $\exp^{-\pi^2/4\theta_c^2}$ in the denominator, assuming $\pi \gg \theta_c)$. 

The first term in the integrand dominates  the second by several orders of magnitude. Therefore, the major contribution to the integral is when the first term, $\frac{\sin{(\theta)}}{(1-\beta(\theta) \cos{(\theta)}^3}$, is at its maximum. This happens at $\theta \sim 1/\Gamma(\theta)$.  For large values of $\Gamma$, therefore the first term at the peak can be approximated as $\frac{1/\Gamma_0}{(1-\beta_0)^3} \sim 8 \Gamma_0^5$.  

Considering contribution only from the $\theta_{\rm max}$ where the first term peaks, $\int d\theta \cal{F}(\theta)$ can be approximated as $\theta_{\rm max} \cal{F}(\theta_{\rm max})$. This reduces the integrand to $\frac{1}{\Gamma_0} \, \Gamma_0^5 \, \frac{1}{\Gamma_0^4}$, leading to the final result that $E_{\rm iso} (0)  \sim \frac{E_{\rm tot,\gamma}}{\theta_c^2}$. In figure-\ref{fig-prompt}, we can see that both top-hat and Gaussian jets result in the same fluence at $\theta_v = 0$.
%%%%%%%%%%%%%%%%%%%%%%%%%%%%%%%%%%%%%%%%%%%%%%%%%%

% Don't change these lines
\bsp	% typesetting comment56
\label{lastpage}
\end{document}